\documentclass[aps,prx,reprint,twocolumn,superscriptaddress]{revtex4-2}

\usepackage{amsmath} 
\usepackage{amssymb} 
\usepackage{graphicx}
\usepackage{dcolumn}
\usepackage{bm}
\usepackage[hidelinks]{hyperref}
\usepackage{color} 
\usepackage{siunitx}
\usepackage{comment}
\usepackage[T1]{fontenc}
\usepackage[capitalise]{cleveref}
\crefname{section}{Sec.}{Secs.}
\usepackage{soul,xcolor}

\usepackage[nearskip=0pt,farskip=0pt,captionskip=0pt,caption=false]{subfig}

\newcommand{\phantomsubfloat}[1]{{
    \captionsetup[subfigure]{labelformat=empty}
    ~\\[-1.6em]
    \subfloat[][]{#1}
}}

\crefrangelabelformat{figure}{#3#1#4--#5(\crefstripprefix{#1}{#2}#6}
\crefmultiformat{figure}{Figs.~#2#1\xdef\mycreffirstarg{#1}#3}{ and~#2({\crefstripprefix{\mycreffirstarg}{#1}}#3}{, #2({\crefstripprefix{\mycreffirstarg}{#1}}#3}{ and~#2({\crefstripprefix{\mycreffirstarg}{#1}}#3}

\DeclareMathOperator{\Tr}{Tr}

\newcommand{\Fig}[2][]{\cref{fig:#2}\if #1\empty\else(#1)\fi}
\newcommand{\eq}[1]{Eq.~\eqref{eq:#1}}

\newcommand{\SM}[1]{SM~\cref{sec:#1}}

\begin{document}

\title{Correlating light fields through disordered media across multiple degrees of freedom}

\author{Louisiane Devaud}
\affiliation{Laboratoire Kastler Brossel, ENS-Université PSL, CNRS, Sorbonne Université, Collège de France, 24 rue Lhomond, 75005 Paris, France}

\author{Bernhard Rauer}
\affiliation{Laboratoire Kastler Brossel, ENS-Université PSL, CNRS, Sorbonne Université, Collège de France, 24 rue Lhomond, 75005 Paris, France}

\author{Simon Mauras}
\affiliation{INRIA, FairPlay joint team, Palaiseau, France}

\author{Stefan Rotter}
\affiliation{Institute for Theoretical Physics, Vienna University of Technology (TU Wien), A-1040 Vienna, Austria}

\author{Sylvain Gigan}
\affiliation{Laboratoire Kastler Brossel, ENS-Université PSL, CNRS, Sorbonne Université, Collège de France, 24 rue Lhomond, 75005 Paris, France}

\date{\today}

\begin{abstract}
Speckle patterns are inherent features of coherent light propagation through complex media. As a result of interference, they are sensitive to multiple experimental parameters such as the configuration of disorder or the propagating wavelength. Recent developments in wavefront shaping have made it possible to control speckle pattern statistics and correlations, for example using the concept of the transmission matrix. In this article, we address the problem of correlating scattered fields across multiple degrees of freedom. We highlight the common points between the specific techniques already demonstrated, and we propose a general framework based on the singular value decomposition of a linear combination of multiple transmission matrices. Following analytical predictions, we experimentally illustrate the technique on spectral and temporal correlations, and we show that both the amplitude and the phase of the field correlations can be tuned. Our work opens up new perspectives in speckle correlation manipulation, with potential applications in coherent control.
\end{abstract}

\maketitle

\section{Introduction}

The propagation of coherent light through a scattering medium results in a complex interference pattern generally called ``speckle''~\cite{Goodman2007}.
Speckle patterns are very difficult to predict and are affected by many experimental parameters: the medium itself, the wavelength or angle of illumination, and, in the case of pulsed light, the propagation time \cite{shapiro1986large,andreoli2015deterministic}.
Because of their complexity and apparent randomness, speckle patterns are often perceived as an inconvenience.
Scattering is a major limitation for imaging both through weakly and strongly diffusive media.
Examples of the former are ubiquitous in astronomy~\cite{labeyrie1970attainment}, and those of the latter in bio-medical imaging~\cite{park2018perspective}.

To overcome difficulties due to scattering, present-day techniques frequently take advantage of correlations such as in the so-called ``optical memory effect'' (ME), which facilitate image reconstruction~\cite{takasaki2014phase,schott2015characterization,yilmaz2015speckle,katz2014non} such as in speckle scanning microscopy~\cite{bertolotti2012non,yang2014imaging}.
Resulting from a small perturbation of the speckle pattern~\cite{feng1988correlations,freund1988memory}, the ME is responsible for maintaining the specific shape of a given speckle pattern when tuning an experimental parameter within a small range (memory effect range).
In a disordered system, however, a whole variety of different parameters can be tuned, resulting in many variants of the ME.
The best known ME for scattering media is the tilt-tilt ME \cite{li1994correlation,schott2015characterization}, generalized by~\cite{osnabrugge2017generalized} to tilt-tilt and shift-shift MEs.
But the ME also involves other degrees of freedom as illustrated by the recent observation of spectro-axial MEs~\cite{zhu2020chromato, devaud2021chromato}.
The ME's limited range, however, drastically restricts its applicability.
This range depends on the medium and is inversely proportional to its thickness so that it is often too small to improve imaging through thick scattering media.
Several previous studies have characterized the ME \cite{schott2015characterization}, aiming to extend its range either by input or output light selection~\cite{kadobianskyi2018scattering,chen2019expansion,yilmaz2019angular} or by engineering specific metasurfaces~\cite{jang2018wavefront,arruda2018controlling}.
However, little effort has been made so far to engineer it.

A versatile approach to the study of complex media is to use wavefront shaping tools and techniques.
Using spatial light modulators (SLM), the incoming beam can be shaped to control the propagation of light.
These techniques allowed introducing non-normally occurring properties \cite{dogariu2015electromagnetic} into speckle patterns as tailoring their statistics through iterative algorithms \cite{han2023tailoring,bender2023spectral} or focusing light behind them \cite{vellekoop2007focusing}.
Another approach is to measure the transmission matrix (TM)~\cite{popoff2010measuring} of the system.
The TM makes it possible to determine the input state that focuses light behind the medium~\cite{mosk2012controlling} or to modulate the energy delivery distribution using its singular modes~\cite{kim2012maximal}.
Non-local correlations of the output field are also possible by appropriate filtering of the Fourier components of the TM~\cite{devaud2021speckle}.
Furthermore, transmission control with a TM, initially developed in monochromatic light, can even be extended to pulses~\cite{mounaix2016deterministic,devaud2022temporal}.

\begin{figure*}[t]
    \includegraphics[width=1\textwidth]{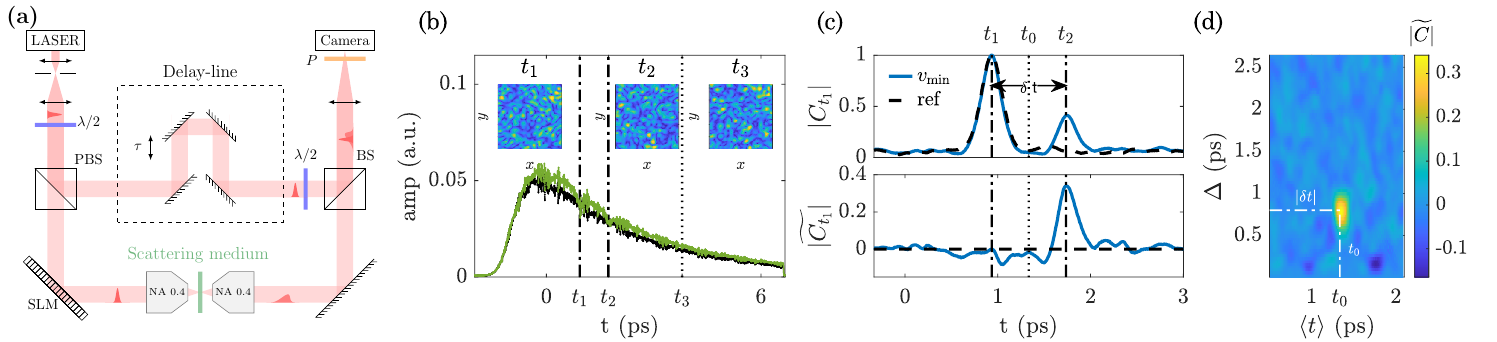}
    \phantomsubfloat{\label{fig:main_effect-a}}
    \phantomsubfloat{\label{fig:main_effect-b}}
    \phantomsubfloat{\label{fig:main_effect-c}}
    \phantomsubfloat{\label{fig:main_effect-d}}
    \caption{Experimental illustration of temporal speckle correlations engineering.
    (a) Schematic of the experimental setup.
    The light delivered by a laser is split into two paths with a half wave plate ($\lambda$/2) and a polarized beam splitter (PBS): (i) signal where the wavefront is modulated by a reflective phase-only SLM and passes through a layer of TiO$_2$ (focused with microscope objectives of 0.4 numerical aperture (NA)); (ii) delay line introducing a delay $\tau$.
    Both paths are recombined with a beam splitter (BS) and their interference, cleaned from the unscattered light with a polarizer (P) is imaged on a CCD camera.
    (b) Averaged temporal amplitude of the field measured behind a thick layer of TiO$_2$ when illuminated by a laser pulse of \SI{100}{\femto \second}.
    The black curve corresponds to the field when a blank pattern is displayed on the SLM whereas the green curve results from the display of $v_{\mathrm{min}}$ (singular vector associated with the smallest singular value of $T_1 - T_2$).
    The time-gated TMs are measured with $N_{\mathrm{SLM}} \approx 620$ modes on the SLM and $N_{\mathrm{CCD}} = 225$ pixels on the CCD-camera at $t_1$ and $t_2$.
    The camera pixels are always binned to have one speckle grain per pixel, except in the images of the field amplitude presented in the insets (for visualization purposes). 
    (c) Top: Temporal evolution of the absolute value of the field correlation with respect to $t_1$ for both the blank SLM pattern (black dashed curve) and for $v_{\mathrm{min}}$. 
    Bottom: subtraction of the reference correlation to highlight only the anomalous correlation increases.
    (d) 2D representation of the correlations to observe simultaneously all the correlations (with subtraction of the reference as in the bottom of~\cref{fig:main_effect-c}).
    For symmetry reasons, only $\Delta = |t - t'|$ i.e., spacing between the two correlated delays, is represented as a function of $ \langle t \rangle = (t + t')/2$, which is the mean value of the two correlated delays.
    A correlation increase is observed for $\Delta = |\delta t| = |t_2 - t_1|$ and  $\langle t \rangle = t_0 = (t_2 + t_1)/2$.
    Data are averaged over 4 realizations of the medium.
    }
    \label{fig:main_effect}
\end{figure*}

Recently, wavefront shaping has been used to create adaptable and reconfigurable MEs.
To our knowledge, three techniques have been presented for this purpose.
Related to the conventional tilt-tilt ME, the authors of~\cite{yilmaz2021customizing} construct an operator to customize the angular memory effect.
In \cite{fan2005principal,carpenter2015observation,xiong2016spatiotemporal,xiong2017principal}, eigenmodes of the Wigner-Smith operator are used to create speckle patterns resistant to frequency variations~\cite{ambichl2017super} -- a manner of frequency ME.
Finally in~\cite{pai2021scattering}, the authors manage to relate the output fields when light is propagating through air or a scattering medium.
One thing all of the techniques have in common is that they rely on a specific TM-based-operator which combines the information for two propagation schemes (i.e., angle, wavelength, and disorder, respectively).
They then use the eigenvectors of this operator as wavefront inputs.
However, so far this approach is limited to a single type of correlation and two output fields.
For the imaging applications based on ME mentioned above, an extension of the current control over the scattered field would improve the speed and accuracy of image reconstruction.

In this article, we address the lack of a robust and unified method for inducing speckle correlations behind a complex medium.
We present a general method for on-demand control/modulation of correlations between any number of speckle patterns.
To do so, we take advantage of the singular value decomposition (SVD) of a linear combination of different TMs.
We first present and apply this technique to the hitherto unexplored temporal domain by correlating the fields for two distinct delays in a pulse (\cref{sec:effect_evidence}).
We present an analytical model that allows us to predict the magnitude of the correlation.
We highlight its correspondence with experimental data in a monochromatic scheme to illustrate the generality of the method (\cref{sec:theory}).
Moreover, taking advantage of the Fourier relation between time and frequency, we investigate periodic correlation in the temporal (spectral) domain that arises when generating the input state with information from spectral (temporal) measurements (\cref{sec:cross_effect}).
Finally, we generalize this effect by presenting the correlations obtained when mixing spectral and temporal information by calculating the input state from the singular value decomposition of the sum of a monochromatic TM and a time-gated TM.

\section{Experimental evidence with time-gated transmission matrices}
\label{sec:effect_evidence}

\subsection{Concept}

Let us start by considering the following problem: one measures two transmission matrices $T_1$ and $T_2$ associated with two different experimental conditions.
How should one shape the input field such that in both cases it leads to the same output?
In a suitable mode basis this output field is described by a vector that results from a corresponding input field vector X:
\begin{equation}
T_1 X =  T_2 X.
    \label{eq:goal}
\end{equation}
We consider~\eq{goal} as the problem of finding a vector in the kernel of $T_1 - T_2$.
When the matrices $T_1$ and $T_2$ are unitary, this is equivalent to finding an eigenvector of $T_1 T_2^\dagger$, as studied in~\cite{pai2021scattering}.
However, in the general case (non-unitary), our method assures that we will find the solution if it exists.
We compute the kernel of $T_1 - T_2$ through its singular value decomposition, which even allows us to deal with rectangular matrices, and yields two output fields that are perfectly correlated (corresponding to a correlation value of 1, see~\cref{eq:correlation_formula}).

\subsection{Experimental implementation}

We first test this hypothesis to generate temporal speckle correlations.
To this end, we send short pulses (\SI{100}{\femto \second}) to a scattering medium (TiO$_2$ layer suspended on a glass coverslip) and measure the intensity on the sample output plane with a CCD camera (see~~\cref{fig:main_effect-a}).
To extract the scattered and elongated transmitted pulse, part of the incoming pulse is diverted to a delay-line that we interfere with the transmitted light following the approach of \cite{miranda2014spatiotemporal,lepetit1995linear}.
By varying the length of the delay-line, we measure the interference for specific delays on the camera and apply filtering to access the temporal field information.
Using this time-gated experimental setup (the details of which are presented in~\SM{SM_setup}) together with a SLM to control the light impinging on the scattering medium, we can measure time-gated TMs and obtain temporal control over the transmission of an incoming pulse of light~\cite{devaud2022temporal}.
The TM measurement is performed following the approach of \cite{popoff2010measuring}, where the camera images are binned such that one speckle grain (defined as the width of the autocorrelation of the amplitude speckle) corresponds to one output mode.
The fields then used for the analysis are binned in the same manner, with no significant difference with respect to non-binned fields, see~\SM{SM_spatial_effect}.

\subsection{Results}

Correspondingly, we measure two time-gated TMs, $T_1$ and $T_2$, at delays of $t_1 =$\SI{0.9}{\pico \second} and $t_2 =$\SI{1.7}{\pico \second}, respectively.
Once the TMs have been measured and normalized with respect to their intensity, we calculate their difference (the normalization allows the two TMs to have the same weight whatever their measured delay in the pulse) and calculate the singular value decomposition of $T_1 - T_2$.
We then select the singular vector associated with the smallest non-zero singular value (corresponding to a normalized value $\tilde{s} = s/\sqrt{\langle s^2\rangle}$ of 0.1), denoted by $v_{\mathrm{min}}$ and display its phase on the SLM to modulate the incoming field.
We extract the temporal evolution of the pulse for this modulation scheme.
The spatially averaged amplitude of the recorded field is shown in~\cref{fig:main_effect-b}.
To study the relationship between the recorded electric fields $E(t)$, we correlate the fields measured for all delays with the field measured at $t_1$ and denote the correlation obtained $C_{t_1}$.
We use the correlation function
\begin{equation}
    C(E_i, E_j) = \frac{E_i^{\dagger}.E_j}{\sqrt{E_i^{\dagger}.E_i. E_j^{\dagger}.E_j}},
    \label{eq:correlation_formula}
\end{equation}
where $E_{i} \equiv E(t_i)$ are complex fields (to get $C_{t_1}$, $t_i = t_1$ and $t_j$ varies).
We plot in~\Fig[c, top]{main_effect} the evolution of the absolute value of $C_{t_1}$.
For the plane wave input (black dashed line), the value reaches 1 for $t = t_1$ and decreases to reach 0 for a delay longer than the temporal speckle grain width.
However, for the shaped input $v_{\mathrm{min}}$, another peak is observed at $t_2$ indicating increased correlation.
To highlight the characteristic of $v_{\mathrm{min}}$,~\Fig[c, bottom]{main_effect} represents the excess correlation of $C_{t_1}$ obtained by subtracting the reference value.
To provide a global visualization of the correlations between all delays $t$, $t'$ we plot in~\cref{fig:main_effect-d} a 2D excess correlation graph (simply named 2D correlation graph in the following) with axis $\left ( \langle t \rangle = (t + t')/2, \Delta = |t - t'| \right )$.
The normalized excess correlation value is given by a color scale and its position in the 2D graph indicates the relative and absolute values between $t$ and $t'$.
Only one correlation peak is visible at the expected position (i.e., $t_0 = (t_1 + t_2)/2 = \SI{1.3}{\pico \second}$ and $\delta t = |t_2 - t_1| = \SI{0.8}{\pico \second}$) confirming that only the correlation between these two times is enhanced.
Note that, the correlation between $t_1$ and $t_2$ is lower than 1, despite what one could expect from~\eq{goal}.
This is due to the experimental limitation of phase-only control of the SLM (see~\cref{fig:phase_impact-a}).

\section{Theoretical expectations and experimental verification}
\label{sec:theory}

\begin{figure}[t]
    \includegraphics[width=1\columnwidth]{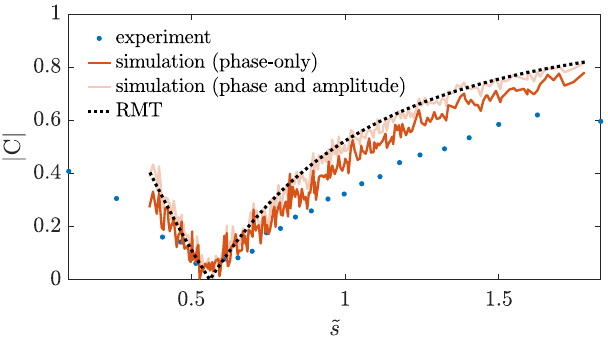}
    \caption{
    Comparison of the absolute value of the correlations when displaying on the SLM different singular vectors for experimental data (for TMs of size $N_{\mathrm{CCD}} = 225$ and $N_{\mathrm{SLM}} \approx 680$), simulations and analytical predictions.
    All correlations are plotted as a function of the normalized singular values $\tilde{s}$.
    The experimental data are represented by the blue dots.
    The analytical prediction of~\cref{eq:corr_value} using the experimental TMs and random matrix theory (RMT) are shown with the black dotted line and the simulations with the orange lines.
    Experimental data are averaged over 4 realizations of the medium.
    }
    \label{fig:theory}
\end{figure}

The above experimental results are consistent with the search for the kernel of $T_1 - T_2$.
However, output fields of singular vectors associated with small singular values have reduced transmission and are more subjected to noise.
For our approach to be relevant for applications (e.g., imaging), it needs to combine high transmission and high correlation.
We thus consider the field correlation obtained by modulating the incoming wavefront with singular vectors associated with singular values larger than zero and especially the largest one, $v_{\rm max}$.
Significant correlation values are also observed in this case, as visible in~\cref{fig:theory}.
Thus, to understand the observations, more accurate modeling is needed.
In the following, we present a general study of the eigenvalue problem for $T_1 + e^{i \alpha}T_2$, where $\alpha$ is an arbitrarily fixed phase (recovering the previous case for $\alpha  = \pi$).
It is important to keep in mind that this mathematical approach corresponds, from a more physical point of view, to determining the wavefront of the input field that favors specific interferences.
In this case, the interference leads, behind the complex media characterized by $T_1$ and $T_2$, to related fields.
This is quite similar to the usual wavefront shaping experiments while here the research is complicated by taking into account not only the propagation information contained in a single TM but in two of them.
The singular value decomposition of $T_1 + e^{i \alpha}T_2$ is equivalent to the eigenproblem of the following matrix product:
\begin{equation}
\begin{aligned}
    \label{eq:svd_expression}
    & (T_1 + e^{i \alpha}T_2)^{\dagger}(T_1 + e^{i \alpha}T_2)   \\
    & = T_1^{\dagger} T_1 + T_2^{\dagger} T_2 +e^{i \alpha} T_1^{\dagger} T_2 + e^{-i \alpha} T_2^{\dagger} T_1.
\end{aligned}
\end{equation}
\begin{figure*}[t]
    \includegraphics[width=1\textwidth]{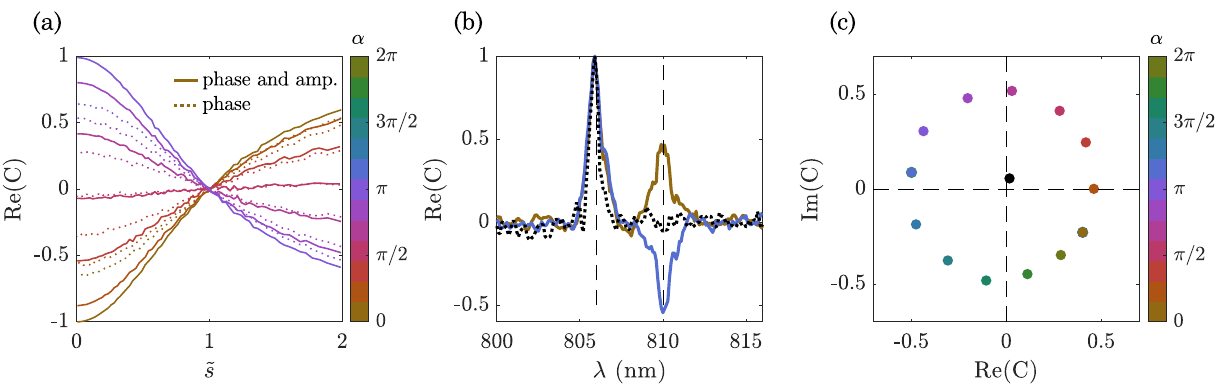}
    \phantomsubfloat{\label{fig:phase_impact-a}}
    \phantomsubfloat{\label{fig:phase_impact-b}}
    \phantomsubfloat{\label{fig:phase_impact-c}}
    \caption{Impact of the relative phase $\alpha$ between the TMs on the field correlations.
    (a) Simulation using $1024 \times 1024$ random matrices.
    Plot of the real part of the correlation for many values of $\alpha$ for phase and amplitude control (solid line) and phase-only control (dotted line).
    The data are averaged over 4 matrix realizations.
    (b) Experimental result of two-wavelength correlations for $\alpha=0$ (brown), $\alpha = \pi$ (purple), and blank input pattern (dotted black).
    TMs are measured for $N_{\mathrm{CCD}} = 225$ and $N_{\mathrm{SLM}} \approx 550 $, the input vector is $v_\mathrm{max}$ and the data are averaged over 4 realizations of the medium.
    (c) Imaginary part of the correlations as a function of the real part for a set of values of $\alpha$ for the same experiment as in (b).
    The central black dot corresponds to the blank input pattern.
    }
    \label{fig:phase_impact}
\end{figure*}
Describing TMs as random matrices~\cite{popoff2010measuring}, we derive in~\SM{SM_analytical_formula} based on \cref{eq:svd_expression} the expected correlation value as a function of the singular values $s$ ($X_s$ is the singular vector associated to the singular value s),
\begin{equation}
    \label{eq:corr_value}
    C(T_1X_s, T_2X_s)  = e^{-i\alpha} \cdot \frac{\tilde s^2-\gamma}{\tilde s^2+ \gamma},
\end{equation}
where $\gamma \equiv N_{\mathrm{CCD}}/N_{\mathrm{SLM}}$ is defined as the ratio between the degree of freedom and the number of controlled modes.
We begin by comparing the analytical model with simulations (including the experimental phase-only constraint) and experiments in~\cref{fig:theory} for no dephasing between the TMs.
We set $\alpha = 0$ and plot the absolute value of the correlation.
The experimental data (blue dots) correspond well to the analytical predictions (black dotted line) as well as to the simulations (orange solid line).
(For a rapid comparison, it is easy to verify that the correlation cancels when $\tilde s = \sqrt{\gamma}$ as predicted by \cref{eq:corr_value}.)
A perfect correspondence is not expected because~\cref{eq:corr_value} is a prediction for full modulation of the wavefront.
The combination of experimental errors and the use of only the singular vector's phase to shape the incoming light leads to a lower measured correlation.

In addition to giving access to the correlation value, \cref{eq:svd_expression} illustrates an important aspect of the method: the existence of two sets of terms.
The first set ($T_1^{\dagger} T_1 + T_2^{\dagger} T_2$) corresponds to a control at the dedicated time-delays $t_1$ and $t_2$.
The second set, consisting of the cross information of the TMs, ($T_1^{\dagger} T_2 + T_2^{\dagger} T_1$) contains the coupling between $T_1$ and $T_2$, which provides the key to control the correlations (see \SM{SM_masks}).
As the induced correlations come from the cross terms only, it is possible to construct the correlation operator directly from this expression.
Note that this is a symmetrized form of the approach of~\cite{pai2021scattering}.

\medskip

So far, we have focused on the absolute value of the correlations, independently of their phase.
However,~\cref{eq:corr_value} highlights the possibility of controlling the phase $\alpha$ between the two fields.
To explore this aspect, we performed field measurements and simulations by varying the relative phase $\alpha$ between two combined TMs.
\Cref{fig:phase_impact-a} presents simulations in agreement with~\cref{eq:corr_value}.
The real part of the correlation as a function of the normalized singular values is plotted for different values of the parameter $\alpha$.
For $\alpha = 0$ or $\pi$ the correlation is real (with opposite signs) and therefore the real part is extreme, for intermediate values an imaginary part exists.
The possibility of reaching a correlated or anti-correlated field is experimentally illustrated in~\cref{fig:phase_impact-b} where two monochromatic TMs are measured at $\lambda_1 = \SI{806}{\nano \metre}$ and $\lambda_2 = \SI{810}{\nano \metre}$.
They are combined with $\alpha$ varying between $0$ and $2\pi$.
For all values of $\alpha$, the SVD is calculated and the phase of $v_\mathrm{max}$ is displayed on the SLM to shape the input beam.
The wavelength is smoothly tuned from \SI{800}{\nano \metre} to \SI{816}{\nano \metre} while recording the field.
The real part of the correlation with the field measured at \SI{806}{\nano \metre} is plotted as a function of $\lambda$.
We observe a correlation when $\alpha = 0$ (brown solid line) and an anti-correlation for $\alpha = \pi$ (blue solid line).
All intermediate values are plotted in the complex plane of~\cref{fig:phase_impact-c} and compared to an input plane wave (black dot).
We have thus demonstrated that the SVD approach allows us to control the amplitude of the correlation by selecting the singular vector according to the associated singular value (radius of the circle in~\cref{fig:phase_impact-c}) and the relative phase $\alpha$ of the fields by fixing the relative phase of the TMs (position on the circle in~\cref{fig:phase_impact-c}) for a given parameter of interest (time-delay, wavelength etc.).

\medskip

\begin{figure}[t]
    \includegraphics[width=1\columnwidth]{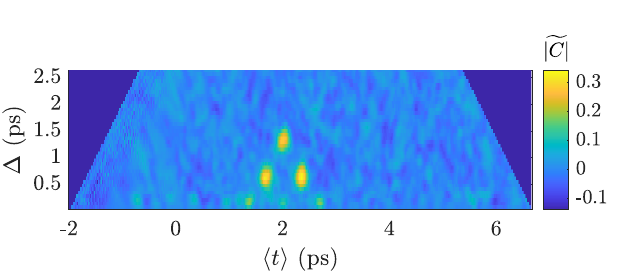}
    \caption{
    Field correlation for three distinct delays in the pulse.
    Measurement of the TMs, calculation of the SVD of ($T_1$ + $T_2$ + $T_3$) and measurement of the temporal field correlation displayed with a 2D correlation plot.
    The data are averaged over 4 realizations of the medium.
    }
    \label{fig:three_times}
\end{figure}

A practical aspect of the above SVD approach is the symmetric roles that the two TMs naturally play.
Based on this symmetrized operator form, an extension to more than two field correlations is straightforward.
To illustrate this point, we measured three time-gated TMs (measured for delays of \SIlist{1.3;2;2.7}{\pico \second}), sum them and calculate the SVD of the new matrix $T_1+T_2+T_3$.
The 2D correlation plot corresponding to the sending of the singular vector associated with the highest singular value ($v_\mathrm{max}$) is presented~\cref{fig:three_times}.
One can observe three correlation peaks (corresponding to the number of combinations of two TMs in a set of three) located at ($t_0 = (t_1+t_2)/ 2 = \SI{1.65}{\pico \second}$, $\delta t = t_2-t_1 =  \SI{0.7}{\pico \second}$), ($t_0 = (t_1+t_3)/ 2 = \SI{2}{\pico \second}$, $\delta t = t_3-t_1 = \SI{1.4}{\pico \second}$) and ($t_0 = (t_2+t_3)/2 = \SI{2.3}{\pico \second}$, $\delta t = t_3-t_2 =  \SI{0.7}{\pico \second}$).
Slight increases are also visible for near-zero $|\delta t|$ values, resulting from an intensity-induced artefact detailed in~\SM{SM_impact_intensity}.
Note that increasing the number of correlated fields unsurprisingly decreases the degree of correlations achieved.
The derivation of the scaling of the correlation with the number of TMs is presented in~\SM{SM_scaling_N}.

It should be noted that the experiments are either performed with pulsed light correlating different delays (\cref{fig:main_effect}, \cref{fig:theory} and \cref{fig:three_times}) or monochromatic light taking the wavelength as a variable parameter (\cref{fig:phase_impact-b,fig:phase_impact-c}) to illustrate the claimed versatility of the approach. 
Furthermore, additional experiments are presented in~\SM{SM_spatial_effect}, showing that fields can also be correlated after propagation through different disorders, generalizing~\cite{pai2021scattering} for any pairs of scattering media.
In the following, we show how to further customize the correlation of the fields.

\section{Cross-effects}
\label{sec:cross_effect}

\begin{figure*}[t]
    \includegraphics[width=1\textwidth]{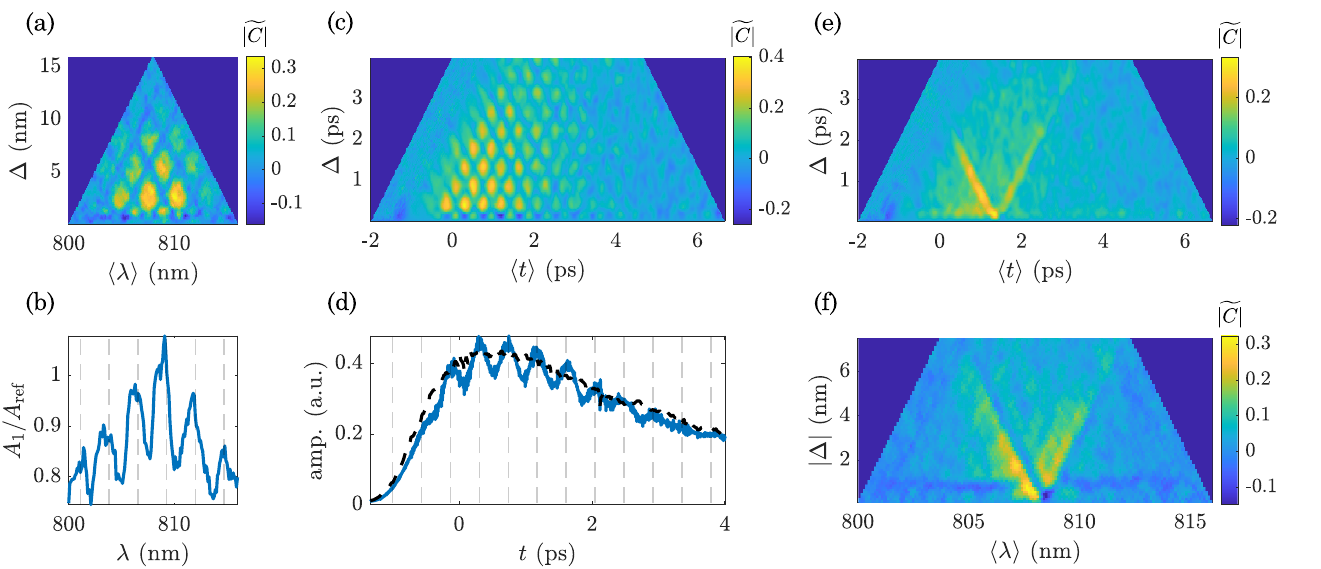}
    \phantomsubfloat{\label{fig:cross_effect-a}}
    \phantomsubfloat{\label{fig:cross_effect-b}}
    \phantomsubfloat{\label{fig:cross_effect-c}}
    \phantomsubfloat{\label{fig:cross_effect-d}}
    \phantomsubfloat{\label{fig:cross_effect-e}}
    \phantomsubfloat{\label{fig:cross_effect-f}}
    \caption{
    Experimental observation of cross-effects.
    (a,b) Time to frequency: measurement of two time-gated TMs at $t_1 = \SI{0.9}{\pico \second}$ and $t_2 = \SI{1.7}{\pico \second}$ for observing the spectral variations when displaying $v_{\mathrm{max}}$ and tuning the wavelength.
    The expected period of the modulation is $\delta \lambda = \lambda_0^2/(c \delta t) = \SI{2.7}{\nano \metre}$.
    (a) 2D correlation plot where (b) represents the amplitude modulation $A_1$ of the singular vector divided by the amplitude $A_{\mathrm{ref}}$ when sending a blank SLM pattern.
    The grey vertical dashed lines illustrate the expected spacing of the peaks.
    (c,d) Frequency to time: measurement of two monochromatic TMs at $\lambda_1 = \SI{805.5}{\nano \metre}$ and $\lambda_2 = \SI{810.5}{\nano \metre}$ and for temporal field control when the sample is illuminated by pulses.
    The expected period of the modulation is $\delta t = \lambda_0^2/(c \delta \lambda) = \SI{0.4}{\pico \second}$ (see vertical dashed lines).
    (c) 2D correlation plot where (d) represents the amplitude modulation (the black dotted line corresponds to the reference amplitude).
    In both cases (frequency to time and time to frequency) the TMs are measured with $N_{\mathrm{SLM}} \approx 2180$ and $N_{\mathrm{CCD}} = 225 $ with the data being averaged over 4 realizations of the medium.
    (e,f) Extension of the correlations by varying different parameters.
    Two TMs with $N_{\mathrm{SLM}} \approx 2220$ and $N_{\mathrm{CCD}}  = 225$ are measured, one is time-gated and one is acquired in monochromatic illumination.
    2D correlation plot for 
    (e) a temporal scan and
    (f) a monochromatic scan when displaying $v_{\mathrm{max}}$.
    The effect being weaker than for the observations in (a) and (b), the data presented are averaged over 8 realizations of the medium.
    }
    \label{fig:cross_effect}
\end{figure*}

We now take advantage of the particular link between time and frequency, which are conjugate quantities.
The idea is to create spectral (temporal) speckle correlations merely using temporal (spectral) information.
We expect such an approach to be useful from the imaging perspective when one domain is easier to access or faster to characterize than the conjugate one.
The information is obtained here by measuring the monochromatic or time-gated TMs and by calculating the input state of interest as described above.
However, we no longer measure the field directly by varying the wavelength or the delay, but we change the laser settings so that we measure the evolution of the field for the conjugate quantity.
In the case of operating from time (characterization of the TMs) to frequency (measurement of the correlations) the procedure is as follows: (i) measure the two time-gated TMs at delays $t_1$ and $t_2$; (ii) compute the SVD of their sum; (iii) modulate the field according to the singular vector of interest ($v_{\mathrm{max}}$); (iv) change the laser settings and measure the field and correlations varying the illumination wavelength.
The experimental results for both cases (i.e., frequency to time and time to frequency) are presented in~\cref{fig:cross_effect}.
The application of beam shaping results in a periodic modulation of the field correlations, well observable in terms of the checkerboard appearing in the 2D correlation plot presented in ~\cref{fig:cross_effect-a,fig:cross_effect-c}.
The period of the modulation is related to the spacing of the frequencies (delays) of the TMs and is given by $\delta \lambda = \lambda_0^2/(c \delta t)$ ($\delta t = \lambda_0^2/(c \delta \lambda)$).
These correlations of the fields are accompanied by amplitude peaks of the same period.
The application of beam shaping results in a periodic modulation of the field amplitude, as can be seen in~\cref{fig:cross_effect-b,fig:cross_effect-d}.
Note that the correlation checkerboard pattern is the generalization to many correlated delays of the peaks observed in the three delays experiment presented in~\cref{fig:three_times} but with the advantage of retaining a high correlation value.
A very interesting aspect of the generation of multiple temporal correlations using frequency measurements is therefore both the drastic reduction of the measurement time and the fact that we retain a high correlation value where a sum of several TMs would have led to its non-negligible reduction.
The latter advantage is also valid for the reciprocal experiment.
The effect of the relative phase $\alpha$ between the TMs for the cross experiment is discussed in~\SM{SM_phase_cross_effect}.

To further extend the possibilities of this technique, we present in the following experiments where we relax the constraint of investigating invariant modes of only one type of ME at a time.
The combined TMs can now be measured in different experimental configurations (not only varying a parameter in one given configuration).
We choose here to combine a monochromatic TM and a time-gated TM.
The monochromatic TM is measured at $\lambda$ = \SI{808}{\nano \meter} and the time-gated TM is measured at a \SI{1.3}{\pico \second} delay in the pulse for the same central wavelength.
Their sum is performed and the phase of the leading singular vector is displayed on the SLM to modulate the incident beam.
Field correlations are extracted and presented in~\cref{fig:cross_effect-e,fig:cross_effect-f} from a pulse scan (e) and a monochromatic scan (f).
In both cases, in the 2D correlation plot, correlation lines are observed.
We observe a ``V-shaped'' correlation whose tip lies at $t_1$ (resp. $\lambda_1$) for $\delta t = 0$ (resp. $\delta \lambda = 0$).
This particular correlation indicates that all speckles at different delays (resp. frequencies) are all partially correlated with the speckle at $t_1$.
Indeed, in this experiment, as for the cross effect, the correlations are spectrally and temporally delocalized.
Here, the sum of the two TMs relates the field information for wavelength $\lambda_1$ and delay $t_1$.
The interpretation is therefore as follows: for all delays, a part of the speckle, with wavelength $\lambda_1$, is correlated to the speckle at $t_1$.
The measured slope of 2 is expected because the horizontal axis shows $t_0$ instead of the delay itself.
This result is particularly interesting because it means that a spectrally selective correlation is
feasible in the pulse.

Extending the field correlation with the SVD to any type of TM makes our approach very practical and promising for future applications.
For example, one could consider combining a TM with a reflection matrix.
Such a choice of matrices provides us with a new experimental tool to investigate how the electric field gets distributed and eventually stored inside a complex medium~\cite{choi2011transmission} by constraining the reflected and transmitted fields to be the same.

\section{Discussion and conclusion}
\label{sec:discuss_other_operators}

We present here a very general method to correlate two or more fields behind a scattering medium with the knowledge of TMs.
Our approach relies on the natural mixing of information that occurs in the SVD of a linear combination of TMs and on the sorting of the singular vectors by the corresponding singular values.
We show that the resulting correlation control is broadly applicable, grants full control of the real and imaginary parts of the correlation, and is determined by the distribution of singular values.
With the modulation of phase and amplitude, even a perfect correlation (or anti-correlation) is possible.
Due to its generality, our approach applies to different illumination configurations such as pulsed light or different wavelengths in monochromatic illumination; also extensions to other experimental platforms such as multi-mode optical fibers~\cite{cao2023controlling} can be considered.
Moreover, our approach is not limited to classical sources since all the techniques used here are also applicable to single photons~\cite{defienne2016two}.
For example, maintaining and engineering correlations between entangled photons, usually scrambled in scattering media, would be extremely valuable for understanding fundamental questions or for developing quantum technologies \cite{safadi2023coherent}.
Finally, because of the many analogies and techniques shared with acoustics or microwaves, our approach should not be restricted to optics \cite{rachbauer2024find,horodynski2022anti,orazbayev2023wave}.\\
This flexible and wide extension of the range of field correlation paves the way for engineering the memory effect at will, with all the practical implications that this entails.
Most promisingly, the correlations are achievable independently of the natural ME range of the medium -- which is typically very small for thick scattering samples, like the one used here.
Taking advantage of the phase of the correlations may therefore prove to be an asset for imaging biological tissues, for example in the context of structured illumination techniques~\cite{gustafsson2000surpassing}.\\
A central point in the field of scattering through complex media is that wavefront-shaping techniques can turn these media into arbitrary optical elements \cite{vellekoop2007focusing,guan2012polarization,park2012active,redding2014high}.
Here the joint control of the intensity and correlations of the field at the output of the medium goes further by opening the way to coherent control.
Our technique makes it possible to create a source with adjustable scattering properties.
When used with pulsed light, the structured pulse and repetition of the same pattern at fixed intervals are ideal for pump-probe experiments, where the ultrafast switching of a shaped wavefront is highly desirable \cite{strudley2014ultrafast}.
For these latter experiments, it is very interesting to note that the presence of the scattering medium allows the illumination to be modulated at a higher rate than the initial pulse repetition rate.
\vspace{5mm}

\begin{acknowledgments}
This project was funding by the European Research Council under the grant agreement No. 724473 (SMARTIES), the European Union's
Horizon 2020 research and innovation program
under the Marie Sk\l odowska-Curie grant agreement No. 888707 (DEEP3P) and the Austrian Science Fund (FWF) under Project No. P32300 (WAVELAND).
\end{acknowledgments}

\bibliography{Biblio_SVD_corr}


\clearpage
\onecolumngrid

\renewcommand{\thefigure}{S\arabic{figure}}
\renewcommand{\theequation}{S\arabic{equation}}
\renewcommand{\thesection}{S\arabic{section}}
\setcounter{equation}{0}
\setcounter{figure}{0}
\setcounter{section}{0}
\newcommand{\Supps}[1]{Supp.~\ref{sec:#1}}

\begin{center}
  \LARGE
  \textbf{Supplementary Materials}
\end{center}

\section{Experimental setup and measurement techniques}
\label{sec:SM_setup}

\begin{figure}[h!]
    \includegraphics[width=0.5\columnwidth]{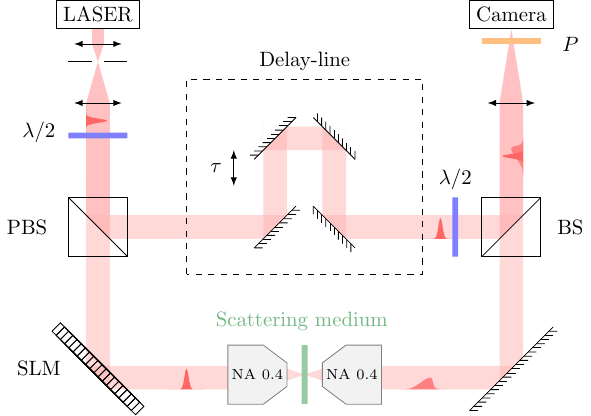}
    \caption{
    Schematic of the experimental setup.
    The light delivered by a laser (MaiTai HP, Spectra Physics) is split into two paths by a polarizing beam splitter (PBS).
    In one path, the wavefront is modulated by a reflective phase-only SLM (HSP512L-1064, Meadowlarks) and passes through a layer of TiO$_2$ (transmittance of $\sim 0.16$, suspended on a glass slide).
    The second path is a controlled delay line.
    The light from both arms is recombined on a beam splitter (BS) which is imaged on a CCD camera (Manta, G-046, Allied Vision).
    A polarizer (P) located before the camera selects the desired polarization.
    }
    \label{fig:SM_setup}
\end{figure}

The experimental setup used for all experiments is presented~\cref{fig:SM_setup}.
The laser, MaiTai HP Spectra Physics, can be used directly in pulsed mode (Gaussian pulse with a full width at half maximum of $\simeq \SI{100}{\femto \second}$) or turned into a tunable monochromatic source.
In both uses, the input light is divided into two beams whose relative intensity is modulated by a half-wave plate.
Each beam follows a different path to allow interferometric measurements.
One arm, whose delay can be adjusted using a delay line, is used as a simple reference in the case of monochromatic light or as a probe pulse in the case of incoming pulsed light.
On the second arm, the wavefront is spatially distorted (and also temporally distorted when sending pulses) by its propagation through a scattering medium.
A phase-only SLM (HSP512L-1064, Meadowlarks) allows the wavefront to be spatially controlled upstream of the scattering medium.
Behind the medium, the two beams are recombined and their interference is imaged on a CCD camera (Manta, G-046, Allied Vision).
A polarizer is placed in front of the camera to select only the multiple scattered light.
The field is recovered from the intensity images using a discrete or continuous phase-stepping holography technique~\cite{devaud2022temporal}.

The diffusing medium used is a layer of TiO$_2$ with a transmittance of $\sim 0.16$, suspended on a glass slide.
The illumination microscope objective focuses the light on the sample and the collection microscope objective associated with a lens images the output surface of the medium on the camera.
Averaging is performed by moving the scattering medium laterally to illuminate different regions of the sample.

To measure TMs we follow the approach of \cite{popoff2010measuring}: we display a set of modes (Hadamard basis, each pixel being either $0$ or $\pi$) on the SLM and record the output field, using a phase-stepping technique, on the camera.
The camera pixels are binned so that one pixel corresponds to one speckle grain (defined as the width of the autocorrelation of the amplitude speckle).
The number of camera pixels (after binning) is denoted $N_{\mathrm{CCD}}$.
On the SLM side, we determine the region of the SLM from which the light passes through the optical setup and is collected.
Only this region of the SLM is modulated and the number of modes effectively controlled is denoted $N_{\mathrm{SLM}}$.
Interference occurs with unmodulated light propagating along the reference arm of the setup.
This method is applied to all measurements of TMs regardless of the nature of the incoming light.
For time-gated measurements, using short pulses, the position of the delay-line determines the gated time whose information is measured.
In contrast, for our monochromatic settings, the length of the reference arm is usually not critical.
However, due to the bandwidth of the source (full width at half maximum of $\SI{1}{\nano \metre}$), some precautions must be taken when measuring two time-gated TMs to perform the cross measurement in frequency.
Indeed, in that case, the position of the stage is not fixed by the experimental protocol.
It is important, when scanning the frequency, to position the delay line at the value of either of one time-gated TMs.
Otherwise, an additional phase difference occurs for the field measurement and correlation lines for $\delta \lambda$ = cst. appear in the 2D correlation plot.
This effect is not intrinsic to the method we present and can be simply observed when measuring a single time-gated TM displaying one vector of its SVD on the SLM and looking at the field correlation for monochromatic light when varying the wavelength.

\section{Impact of intensity on correlations}
\label{sec:SM_impact_intensity}

\begin{figure}[h!]
    \includegraphics[width=0.7\columnwidth]{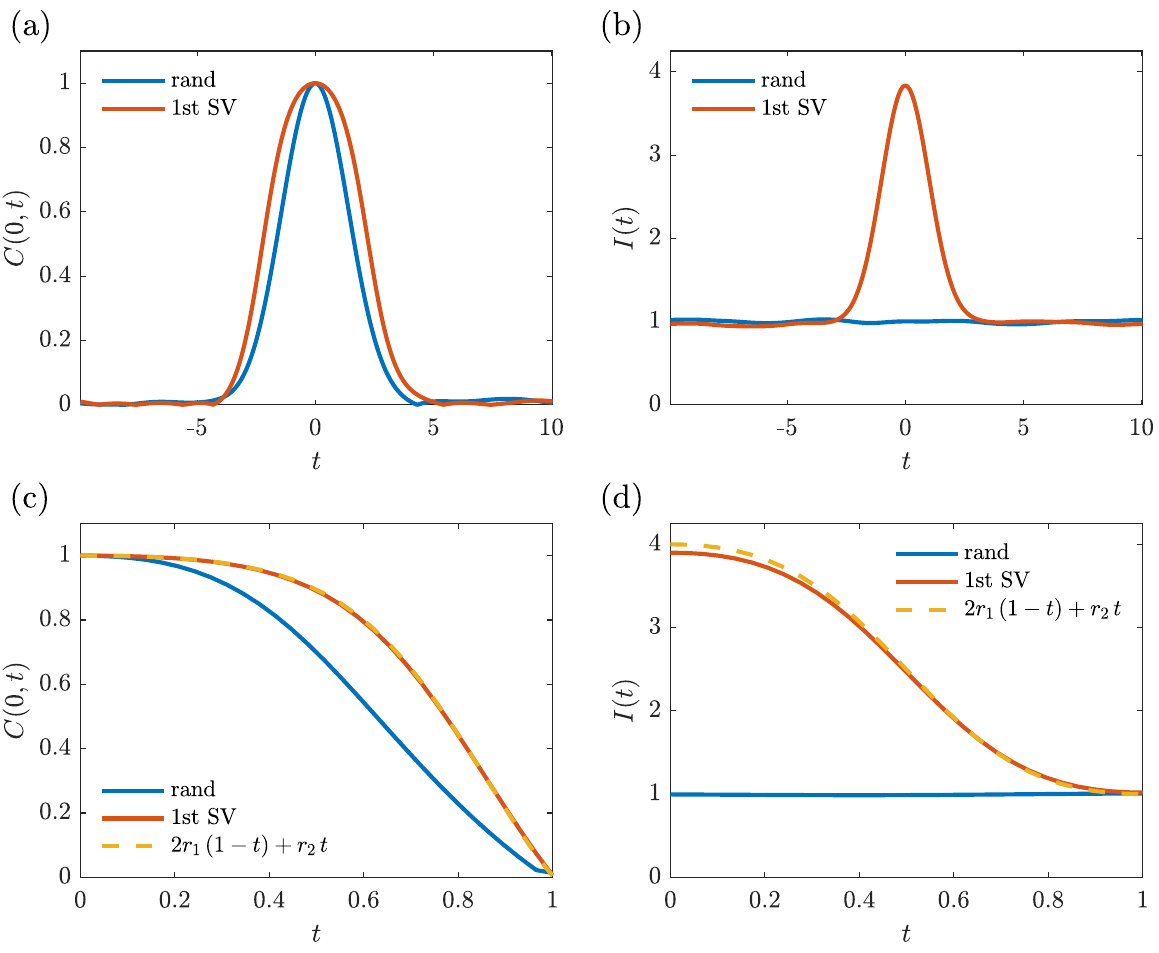}
    \phantomsubfloat{\label{fig:SM_int_corr-a}}
    \phantomsubfloat{\label{fig:SM_int_corr-b}}
    \phantomsubfloat{\label{fig:SM_int_corr-c}}
    \phantomsubfloat{\label{fig:SM_int_corr-d}}
    \caption{
    Correlations induced by intensity variations.
    The TMs are of size $256 \times 256$ and sampled every 0.1-time unit (a,b) or 0.03 (c,d).
    (a) Field correlations between $t=0$ and other times for a random input (blue) and the singular vector associated with the largest singular value of the TM measured at $t=0$ (1st SVD, red).
    (b) Total intensity for the same data as in (a).
    (c) Field correlations for a random input (blue), the singular vector associated with the largest singular value (red) and an artificial random input having the same time evolution as the TMs (dashed yellow).
    (d) Intensity for the same data as in (c).
    The data are averaged over 20 realisations of the disorder (a,b) and 10 realisations (c,d).
    }
    \label{fig:SM_int_corr}
\end{figure}

As visible in the correlation formula given in~\cref{eq:correlation_formula} of the main text, the correlated fields are normalized to avoid intensity-induced artefacts.
However, this does not fully prevent intensity changes from influencing the correlation measure.
For example, imagine a system with a TM that continuously changes in time.
For a random input, the output state of this TM will show the same temporal correlations as the matrix itself.
Yet, if we calculate the output generated by the first singular vector of the TM at a given point in time $t_0$ we see that its temporal correlations are wider (\cref{fig:SM_int_corr-a}).
Does this mean that the output of the singular vector is more stable in time?
It does, but the effect is only due to the intensity enhancement associated with the first singular vector (\cref{fig:SM_int_corr-b}).
As the output speckle at this point in time has a much higher intensity it also has a higher contribution at neighbouring times. This is something the normalization in \cref{eq:correlation_formula} does not account for.

To understand that this is solely caused by intensity variations, let us consider an even simpler example:
A TM that is composed of only two components with linearly changing weights $T \propto T_1 (1-t) + T_2 t$.
Also here, a random input decorrelates faster than the first singular vector of $T_1$ (\cref{fig:SM_int_corr-c}).
However, if we construct an artificial random output with the temporal evolution $r = 2 r_1 (1-t) + r_2 t$, given that $r_{1,2}$ are random vectors, we observe the same temporal correlations as for the singular vector output.
This shows that simply by matching the intensity evolution of the singular vector output state we could reproduce the same correlation feature (\cref{fig:SM_int_corr-d}).
Therefore, here the enhanced correlations are only due to intensity differences and not inherent to the SVD.

In the measurements, this effect is visible for small $\delta t$ or $\delta \lambda$, for example, in \cref{fig:three_times} or \cref{fig:SM_masks}.
However, it only acts on scales of the temporal decorrelation length and does not affect the long-range correlations investigated in this work.
This can be seen in  \cref{fig:main_effect} of the main text.
There, field correlations are observed between two different delay times in the pulse that show a slightly reduced intensity.
This leads to a decreased width of the speckle auto-correlations around these times, i.e., the inverse of the effect explained in \cref{fig:SM_int_corr}.
In the 2D representation of the correlations presented in \cref{fig:main_effect-c}, this manifests as two dark spots around $t_0 = t_{1,2}$ but does not affect the SVD-induced correlations at larger $\delta t$.

Note that, the intensity artefacts described here occur in complex media where the speckle pattern and the transmitted intensity vary.
In the case of multimode fibres, where the transmission is generally constant, we do not expect these effects to play a significant role~\cite{xiong2017principal}.

\section{Evolution of the effect with the temporal delay between TMs}
\label{sec:SM_diff_times}

\begin{figure}[h!]
    \includegraphics[width=1\columnwidth]{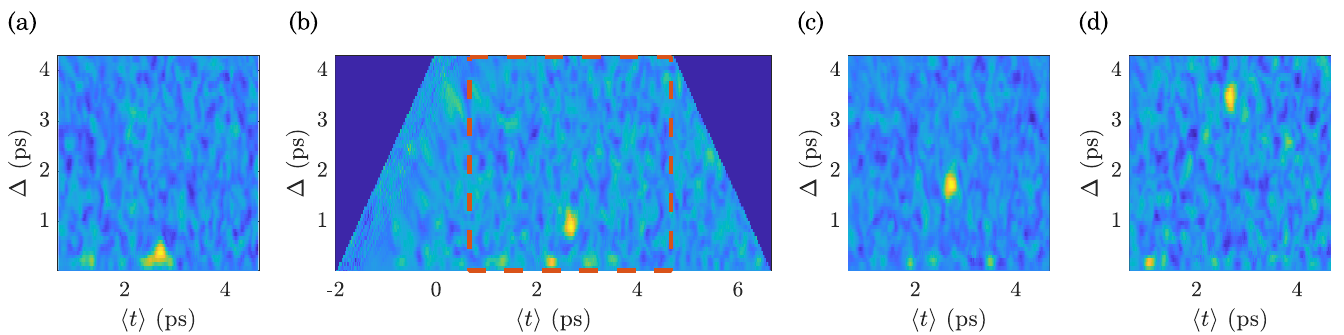}
    \phantomsubfloat{\label{fig:SM_diff_times-a}}
    \phantomsubfloat{\label{fig:SM_diff_times-b}}
    \phantomsubfloat{\label{fig:SM_diff_times-c}}
    \phantomsubfloat{\label{fig:SM_diff_times-d}}
    \caption{
    Evolution of the appearance of the correlation plot when the delay time between the two measured time-gated TMs ($N_{\mathrm{CCD}} = 225$ and $N_{\mathrm{SLM}} \approx 680$) is modified from \SI{0.4}{\pico \second} (a) to \SI{3.2}{\pico \second} (d) while keeping the same central position.
    The correlation peak appears at a fixed $t_0$ but with a $|\delta t|$ that varies with the relative position of the time-gated TMs.
    The full correlation is only plotted in (b) whereas on the other graphs, only the area of the dashed red rectangle is shown.
    The data are not averaged.
    }
    \label{fig:SM_diff_times}
\end{figure}

To illustrate the variation of the position of the peak in the correlation plot, a set of experiments is performed keeping the same central time ($t_0$ = \SI{2.6}{\pico \second}) but varying the delay $\delta t$ between the two time-gated TMs.
The results are presented~\cref{fig:SM_diff_times}.
The correlation peak moves at a fixed $t_0$ along $|\delta t|$.

\section{Difference with masks sum and correlation technique}
\label{sec:SM_masks}

\begin{figure}[h!]
    \includegraphics[width=0.8\columnwidth]{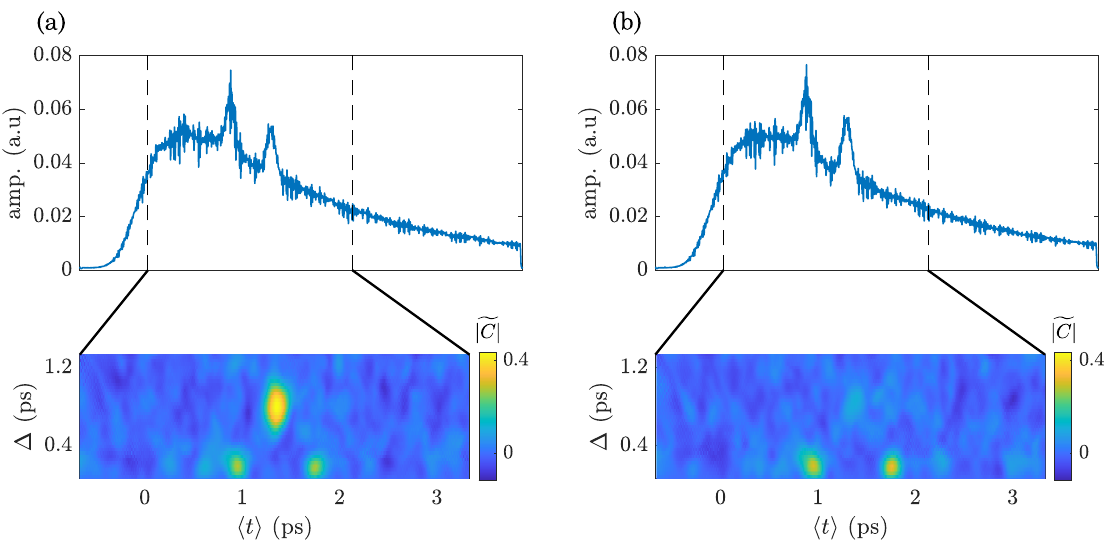}
    \phantomsubfloat{\label{fig:SM_masks-a}}
    \phantomsubfloat{\label{fig:SM_masks-b}}
    \caption{Importance of the SVD terms coupling the two TMs.
    (a) SVD of ($T_1 + T_2$) and (b) coherent sum of the phase masks obtained from the SVD of ($T_1$) and the SVD of ($T_2$) performed separately.
    The top parts represent the pulse shape in both cases.
    The intensity-induced correlations are visible on the 2D correlation plot at the bottom.
    No further correlations are present in the coherent sum case (b) whereas one appears for $|\delta t|$ = \SI{0.8}{\pico \second} in (a).
    TMs are measured $N_{\mathrm{CCD}} = 225$ and $N_{\mathrm{SLM}} \approx 570$.
    Data averaged over 4 realizations of the medium.}
    \label{fig:SM_masks}
\end{figure}

To emphasise the importance of the cross terms in \cref{eq:svd_expression} in the main text, we compared the correlations and pulse shapes of the following two situations.
For \cref{fig:SM_masks-a}, the usual procedure of this work is applied, i.e., we measure two time-gated TMs at two different times in the pulse, add them together and perform the SVD of the sum.
We measure both the correlations and the pulse shape when the first calculated singular vector is displayed on the SLM.
One observes intensity peaks at the position two times of the time-gated TMs visible both on the pulse shape and on the 2D correlation plot (see~\cref{sec:SM_impact_intensity}) for small $|\delta t|$.
There is also a correlation peak corresponding to the correlation between the field at the two time-gated TMs times, thus appearing at $t_0 = (t_1 + t_2)/2 = \SI{1.3}{\pico \second}$ and $|\delta t| = t_2 - t_1 = \SI{0.8}{\pico \second}$.
On \cref{fig:SM_masks-b} the SVD of the two time-gated TMs are performed separately and their 1st singular vectors are summed to create the vector whose phase is displayed on the SLM, as for the multi-time control~\cite{mounaix2016deterministic,devaud2022temporal}.
In the latter case, the intensity enhancement is still present (both in the pulse shape and the 2D correlation plot) but no further field correlation is observable.

In~\cref{fig:main_effect-a} of the main text, due to both the weak control given by the ratio between the degree of freedom and the number of controlled modes $\gamma \equiv N_{\mathrm{CCD}}/N_{\mathrm{SLM}} \simeq 0.36$ and the use of the singular vector associated to the smallest singular value, no clear intensity variation is visible for this specific input state (green curve) compared to the field obtained for a plane wave input (black curve).

\section{Phase impact for the cross effect}
\label{sec:SM_phase_cross_effect}

\begin{figure}[h!]
    \includegraphics[width=0.8\columnwidth]{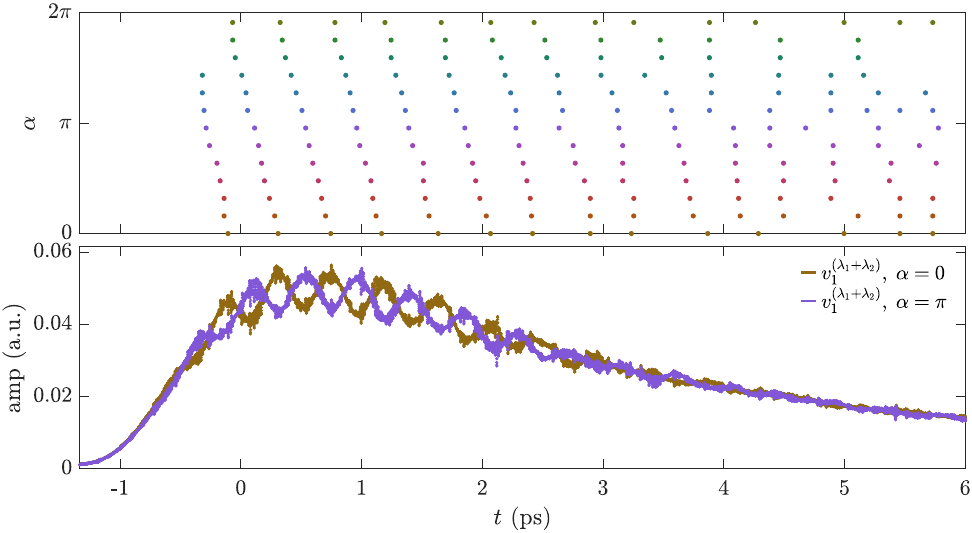}
    \caption{Impact of $\alpha$ on the frequency to time cross effect.
    The bottom plot represents the shape of the pulse for $\alpha$ = 0.
    The top part represents the tracking of the maxima of the pulse while varying $\alpha$: the maxima are shifted.
    The data come from the same experiment as in~\cref{fig:cross_effect}, thus averaged over 4 realizations of the medium.
    }
    \label{fig:SM_phase_cross_effect}
\end{figure}

The effect of the relative phase $\alpha$ between the two TMs has already been studied and the main text (\cref{sec:theory}).
It is presented in the case of two monochromatic TMs in~\cref{fig:phase_impact}.
Here we present its impact on the cross effect.
In~\cref{fig:SM_phase_cross_effect} as in~\cref{fig:cross_effect-c,fig:cross_effect-d} we measure two monochromatic TMs and extract the singular vectors of their sum for different values of $\alpha$.
A pulse scan is performed to extract the field.
Changing $\alpha$ causes the complete correlation plot to be translated along $t_0$.
To visualise this, one can look at the shape of the extracted pulse when performing the frequency-to-time cross effect.
On the top part of~\cref{fig:SM_phase_cross_effect} we track the positions of the maxima for all values of $\alpha$.
For $\alpha$ from 0 to 2$\pi$, all the maxima are shifted by one period. 

\section{Spatial effect}
\label{sec:SM_spatial_effect}

\begin{figure}[h!]
    \includegraphics[width=0.8\columnwidth]{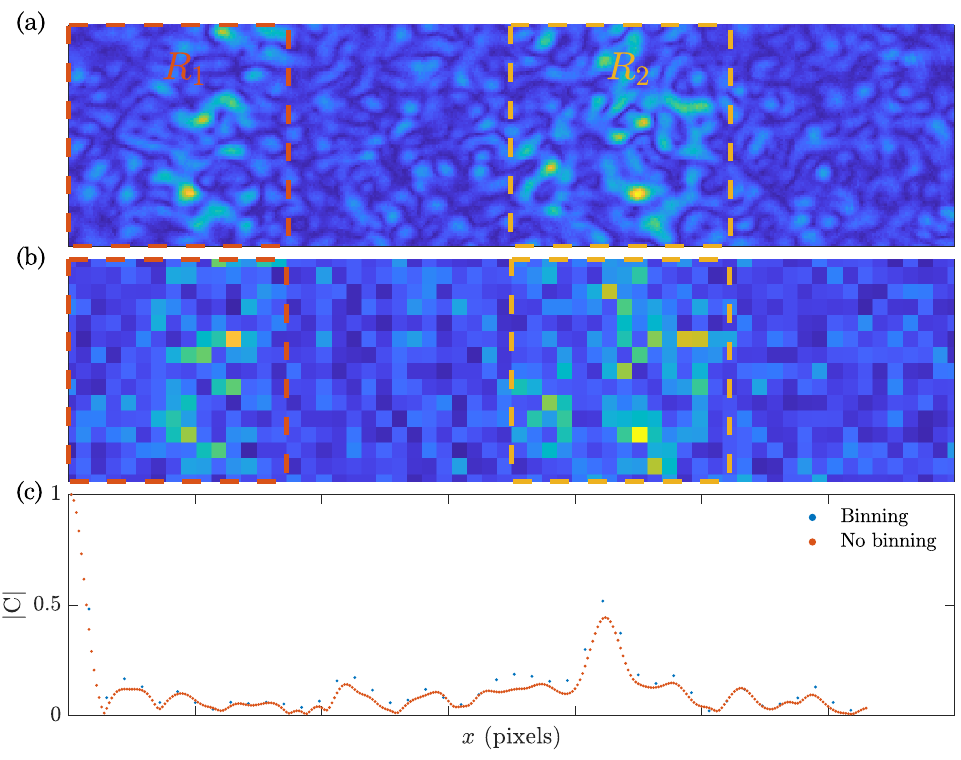}
    \phantomsubfloat{\label{fig:SM_spatial_effect-a}}
    \phantomsubfloat{\label{fig:SM_spatial_effect-b}}
    \phantomsubfloat{\label{fig:SM_spatial_effect-c}}
    \caption{Spatial field correlation
    (a) Unbinned image of the speckle.
    The TMs are measured on the left ($R_1$) and central ($R_2$) parts for $\lambda = \SI{808}{\nano \metre}$, $N_{\mathrm{CCD}} = 225$ and $N_{\mathrm{SLM}} \approx 610$.
    (b) Same speckle image but binned so that one pixel maps one speckle grain.
    (c) Correlation of the left ROI ($R_1$) to all the other positions with a moving window.
    An increase in correlation is observed when the moving window reaches the position where the other TM was measured.
    Data averaged over 4 realizations of the medium.
    }
    \label{fig:SM_spatial_effect}
\end{figure}

As stated in the main text, the method we present is very general and does not depend on the parameter used to vary the TM.
We have worked mainly with time and frequency but other parameters can be used.
To show this explicitly, we present here the same experiment performed for different spatial positions.
We measure two TMs linking the plane of the SLM to two different regions of interest (ROIs) $R_1$ and $R_2$ on the camera.
TMs are, as usual, summed and the phase of the first singular vector obtained is displayed on the SLM.
We extract the field extract for a large ROI  of the camera including the two regions where the TMs were measured.
We display in~\cref{fig:SM_spatial_effect-a,fig:SM_spatial_effect-b} the amplitude of the fields with or without binning of the camera pixels.
We calculate the correlations between $R_1$ and all other possible ROIs using a moving window, as shown in~\cref{fig:SM_spatial_effect-c}.
An increase in correlation can be observed when the moving window coincides with $R_2$.
The correlation values obtained for the binned and unbinned images are similar.
In the case of the unbinned image, the correlation value is slightly lower, this behaviour is expected as the image contains information about the high k vectors that is not contained in the TM itself.
In~\cref{fig:SM_spatial_effect} the geometry presented is simple because the scan is only performed on one dimension.
However, the result generalises to a two-dimensional scan.

\section{Analytical model and simulations}
\label{sec:SM_model_sim}

\subsection{Analytical formula for the correlation}
\label{sec:SM_analytical_formula}

In this section, we derive a formula for the correlation between the output fields of two random transmission matrices, when sending a singular vector of a linear combination. Formally, let $T_1$ and $T_2$ be $n$-by-$m$ matrices where all coefficients are drawn independently from the same (complex) Gaussian distribution of mean $0$ and standard deviation $\sigma$. Fix two parameters $\alpha_1$ and $\alpha_2$ and define
\[
M = \frac{e^{i\alpha_1}T_1+e^{i\alpha_2}T_2}{\sqrt{2}}
\qquad\text{and}\qquad
\Delta = \frac{e^{i\alpha_1}T_1-e^{i\alpha_2}T_2}{\sqrt{2}}.
\]
An important property is that our Gaussian distribution is invariant to rotations, such as the one transforming $(T_1,T_2)$ into $(M,\Delta)$. More precisely, coefficients of $(T_1,T_2)$ are independent Gaussian variables of mean $0$ and standard deviation $\sigma$, thus coefficients of $(M,\Delta)$ are also independent Gaussian variables of mean $0$ and standard deviation~$\sigma$. Therefore, we draw $M$ which is now a fixed matrix, but $\Delta$ is still random.

\smallbreak
We write the singular value decomposition $M = U\Sigma V^\dagger$, where $U \in \mathbb C^{n\times n}$ and $V \in \mathbb C^{m\times m}$ are unitary matrices and $\Sigma\in\mathbb R_+^{n\times m}$ is diagonal. Let $X \in \mathbb C^m$ and $Y\in \mathbb C^n$ be the right and left singular vectors associated with a singular value $\mu$. More precisely, $\mu$ is the $j$-th coefficient of $\Sigma$ for some $j$, and $X$ and $Y$ are respectively the $j$-th column of $V$ and $U$. In particular, we have
\[M X = \mu Y
\qquad\text{and}\qquad
Y^\dagger M = \mu X^\dagger\]
Recall now the correlation $C(T_1X, T_2X)$ between vectors $T_1X$ and $T_2X$.
\[C(T_1X, T_2X) = \frac{X^\dagger T_1^\dagger T_2 X}{||T_1X||_2\cdot ||T_2X||_2}\]
Rewritting $T_1X$ and $T_2X$ using $M$ and $\Delta$, we obtain
\begin{align*}
\sqrt{2}e^{i\alpha_1} T_1X &= (M+\Delta)X = \mu Y+\Delta X\\
\sqrt{2}e^{i\alpha_2} T_2X &= (M-\Delta)X = \mu Y-\Delta X
\end{align*}
We define the random variable $z = Y^\dagger \Delta X =\sum_{j=1}^n\sum_{k=1}^m  \overline{Y_j} X_k\Delta_{j,k}$. Recall that $\Delta$ is random Gaussian and that $X$ and $Y$ are fixed unit vectors. Therefore $z$ is also Gaussian, and a short calculation shows it has mean 0 and standard deviation $\sigma$.
\begin{align*}
||T_1X||_2^2 = X^\dagger T_1^\dagger T_1 X =
(\mu Y^\dagger+X^\dagger\Delta^\dagger)(\mu Y+ \Delta X)/2 &=
(\mu^2+||\Delta X||_2^2)/2 + \mu\Re(z)\\
||T_2X||_2^2 = X^\dagger T_2^\dagger T_2 X =
(\mu Y^\dagger-X^\dagger\Delta^\dagger)(\mu Y- \Delta X)/2 &=
(\mu^2+||\Delta X||_2^2)/2 - \mu\Re(z)
\end{align*}
We are finally ready to write the correlation between $T_1X$ and $T_2X$.
\begin{align*}
C(T_1X, T_2X) &= \frac{
e^{i(\alpha_1-\alpha_2)}
\left(\mu Y^\dagger + X^\dagger \Delta^\dagger\right)
\left(\mu Y - \Delta X\right)/2
}{
\sqrt{(\mu^2+ ||\Delta X||_2^2)/2 + \mu \Re(z)}
\sqrt{(\mu^2+ ||\Delta X||_2^2)/2 - \mu \Re(z)}
}\\
&=e^{i(\alpha_1-\alpha_2)} \cdot
\frac{\mu^2-||\Delta X||_2^2-2i\Im(z)}{\sqrt{(\mu^2+||\Delta X||_2^2)^2 - 4\mu^2 \Re(z)^2}}
\end{align*}
A short calculation gives $||\Delta X||_2^2 = \sum_{j=1}^n |\sum_{k=1}^m \Delta_{j,k}X_k|^2$. Each inner sum is a Gaussian random variable of mean 0 and standard deviation $\sigma$, thus the expected value of $||\Delta X||_2^2$ is exactly equal to $n\sigma^2$.
Notice that up to this point, all the formulae are exact. We will now approximate random variables by their expected value (which can be formalized with concentration inequalities) and say that $||\Delta X||_2^2 \approx n \sigma^2$. Thus, terms involving $z$ are negligible, and we write
\[
C(T_1X, T_2X) \approx e^{i(\alpha_1-\alpha_2)} \cdot \frac{\mu^2-n\sigma^2}{\mu^2+n\sigma^2}.
\]
To conclude, we define the aspect ratio $\gamma = n/m$. We normalize $\tilde \mu^2 = \mu^2/\langle \mu^2\rangle$, computing the average over $n$ singular values (including zeros if $\gamma > 1$). More precisely $\langle \mu^2 \rangle = \Tr(\Sigma^\dagger \Sigma)/n = \Tr(M^\dagger M)/n = \sum_{j,k} |M_{j,k}|^2/n \approx m\sigma^2$, thus
\begin{equation}
C(T_1X, T_2X) \approx e^{i(\alpha_1-\alpha_2)} \cdot \frac{\tilde \mu^2-\gamma}{\tilde \mu^2+\gamma}
\label{eq:SM_correlation_general}
\end{equation}
Observe that in the formula, both $X$ and $\tilde \mu$ are invariant when multiplying $M$ by a constant (here $\sqrt{2}$). Hence, we can instead compute the singular value decomposition of $e^{i\alpha_1}T_1 + e^{i\alpha_2}T_2$, which exactly corresponds to \cref{eq:corr_value} from the main text when $\alpha_1 = 0$, $\alpha_2 = \alpha$, $n = N_{\mathrm{CCD}}$ and $m= N_{\mathrm{SLM}}$. Finally, Marchenko-Pastur's law says that $\tilde \mu^2$ varies between $(1-\sqrt{\gamma})^2$ and $(1+\sqrt{\gamma})^2$, so correlations vary between $e^{i(\alpha_1-\alpha_2)}\frac{1-2\sqrt{\gamma}}{1-2\sqrt{\gamma}+2\gamma}$ and $e^{i(\alpha_1-\alpha_2)}\frac{1+2\sqrt{\gamma}}{1+2\sqrt{\gamma}+2\gamma}$.

\subsection{Scaling of the correlation with the number of TMs}
\label{sec:SM_scaling_N}

We will now generalize the result of the previous section to linear combinations of $\ell$ random matrices. The main technical insight is that the rotation $(T_1,T_2)\mapsto(M,\Delta)$ used in the previous section to deal with conditional probabilities can be replaced by a unitary transformation $R$.

\smallbreak
For convenience, we start by introducing some notations. For any matrix $X$ of size $a \times b$, we denote $X^{(c,d)}$ its reshaping of size $c\times d$, with $ab=cd$, where coefficients of the matrix are read in row-major order. In particular, $X^{(1,ab)}$ is the vectorization of $X$, that is the row-vector containing coefficients of $X$. For convenience, we will write $X^{(a,b)}$ when we want to emphasize the fact that $X$ has size $a\times b$.

\smallbreak
Let $T_1^{(n,m)}, \dots, T_\ell^{(n,m)}$ be a collection of $\ell$ random matrices of size $n \times m$, whose coefficients are drawn independently from a complex normal distribution of mean $0$ and standard deviation $\sigma$. We define a matrix $T^{(\ell,nm)}$ of size $\ell\times nm$, whose $j$-th row is $T_{j}^{(1,nm)}$. Chose a unitary matrix $R = (r_{j,k})_{1 \leq j,k \leq \ell}$ of size $\ell \times \ell$, that can be thought as a rotation matrix, and let $S = RT$. For every $j$, we define a matrix $S_j^{(n,m)}$ of size $n\times m$ equal to the reshaping of the $j$-th row of $S$. We have the following relations:
\[
\text{for all } 1 \leq j \leq \ell,\qquad
S_j = \sum_{k=1}^\ell r_{j,k} \cdot T_k
\qquad\text{and}\qquad
T_j = \sum_{k=1}^\ell \overline{r_{k,j}} \cdot S_k.
\]
We will now have a closer look at $M = S_1$. Classical results on multivariate normal distributions show that coefficients from $S$ are independent and normally distributed with mean $0$ and standard deviation $\sigma$. Therefore, we draw $M$ which is now a fixed matrix, but $S_2, \dots, S_\ell$ are still random.

\smallbreak  We write the singular value decomposition $M = U \Sigma V^\dagger$ where $U^{(n,n)}$ and $V^{(m,m)}$ are unitary and $\Sigma^{(n,m)}$ is diagonal. Let $U_q^{(n,1)}$ and $V_q^{(m,1)}$ be the $q$-th left and right singular vectors of $M$ associated to the singular value $\mu_q$; that is $U_q$, $V_q$ and $\mu_q$ are respectively the $q$-th columns of $U$ and $V$, and the $q$-th coefficient of $\Sigma$.

\smallbreak

For every $j_1,j_2,q$, we will compute the correlation between $T_{j_1} V_q$ and $T_{j_2} V_q$, for which we need the scalar product
\begin{align*}
  (T_{j_1} V_q)^\dagger (T_{j_2} V_q) =\;&  V_q^\dagger
  \left(\sum_{k_1=1}^\ell r_{k_1,j_1} \cdot S_{k_1}^\dagger\right)
  \left(\sum_{k_2=1}^\ell \overline{r_{k_2,j_2}} \cdot S_{k_2}\right)
  V_q\\
  =\;& r_{1,j_1}\cdot \overline{r_{1,j_2}}
    \cdot V_q^\dagger M^\dagger M V_q
  + \sum_{k=2}^\ell r_{k,j_1}\cdot \overline{r_{k,j_2}}
    \cdot V_q^\dagger S_k^\dagger S_k V_q
  +\sum_{k_1\neq k_2} r_{k_1,j_1} \cdot \overline{r_{k_2,j_2}}
    \cdot V_q^\dagger S_{k_1}^\dagger S_{k_2} V_q
\end{align*}
From the singular value decomposition, we have $V_q^\dagger M^\dagger M V_q = \mu_q^2$. For each $k \neq 1$, the matrix $S_k$ is independent from $S_1$, and thus from the vector $V_q$. Hence, the same argument as in the previous section shows that $V_q^\dagger S_k^\dagger S_k V_q = ||S_k V_q||_2^2\approx n\sigma^2$. For each $k_1 \neq k_2$ the vectors $S_{k_1}V_q$ and $S_{k_2} V_q$ are independent, and because the dimension is high they are approximately orthogonal, thus $V_q^\dagger S_{k_1}^\dagger S_{k_2} V_q \approx 0$.

\medbreak
To finalize the formula, we need a disjunction on whether or not $j_1$ and $j_2$ are equal. When $j_1 \neq j_2$, we can use the fact that the matrix $R$ is unitary, to write $\sum_{k=2}^\ell r_{k,j_1}\cdot\overline{r_{k,j_2}} = -r_{1,j_1}\cdot\overline{r_{1,j_2}}$. 
\[\forall j_1 \neq j_2,\qquad (T_{j_1} V_q)^\dagger (T_{j_2} V_q) \approx r_{1,j_1}\cdot \overline{r_{1,j_2}}\cdot(\mu_q^2-n\sigma^2)\]
When $j_1=j_2=j$ we have $\sum_{k=2}^\ell r_{k,j_1}\cdot\overline{r_{k,j_2}} = 1-|r_{1,j}|^2$.
\[\forall j,\qquad (T_{j} V_q)^\dagger (T_{j} V_q) \approx |r_{1,j}|^2\cdot(\mu_q^2-n\sigma^2)+n\sigma^2\]
Thus, the correlation formula is
\[\forall j_1 \neq j_2,\qquad C(T_{j_1} V_q, T_{j_2} V_q) \approx
\frac{r_{1,j_1} \cdot \overline{r_{1,j_2}}\cdot(\mu_q^2 - n\sigma^2)}{
\sqrt{|r_{1,j_1}|^2\cdot(\mu_q^2-n\sigma^2)+n\sigma^2}
\sqrt{|r_{1,j_2}|^2\cdot(\mu_q^2-n\sigma^2)+n\sigma^2}
}\]
As in the previous section, we introduce the aspect ratio $\gamma = n/m$ and we normalize $\tilde \mu^2 = \mu^2/\langle \mu^2\rangle \approx \mu^2 / (m\sigma^2)$.
\[\forall j_1 \neq j_2,\qquad C(T_{j_1} V_q, T_{j_2} V_q) \approx
\frac{r_{1,j_1}}{|r_{1,j_2}|}\cdot\frac{\overline{r_{1,j_2}}}{|r_{1,j_1}|}\cdot
\frac{\tilde \mu_q^2-\gamma}{
\sqrt{\tilde \mu_q^2+\gamma\cdot\big(\frac{1}{|r_{1,j_1}|^2}-1\big)}
\sqrt{\tilde \mu_q^2+\gamma\cdot\big(\frac{1}{|r_{1,j_2}|^2}-1\big)}
}
\]
In particular, computing the singular value decomposition of $\sum_{j=1}^\ell e^{i\alpha_j} T_j$ corresponds to coefficients $r_{1,j} =e^{i\alpha_j}/\sqrt{\ell}$ in the first row of $R$, in which case the correlation formula can be simplified into
\begin{equation}
\forall j_1 \neq j_2,\qquad C(T_{j_1} V_q, T_{j_2} V_q) \approx
e^{i(\alpha_{j_1}-\alpha_{j_2})}\cdot\frac{\tilde \mu_q^2-\gamma}{\tilde \mu_q^2+\gamma\cdot\big(\ell-1\big)}
\end{equation}
Using random matrix theory, Marchenko-Pastur law shows that $\tilde \mu_q^2$ ranges from $(1-\sqrt{\gamma})^2$ to $(1+\sqrt{\gamma})^2$, and the correlation ranges from $e^{i(\alpha_{j_1}-\alpha_{j_2})}\cdot\frac{1-2\sqrt{\gamma}}{1-2\sqrt{\gamma}+\ell\gamma}$ to $e^{i(\alpha_{j_1}-\alpha_{j_2})}\cdot\frac{1+2\sqrt{\gamma}}{1+2\sqrt{\gamma}+\ell\gamma}$.

\subsection{Random unitary matrices}

Let us now consider two unitary random matrices $T_1$ and $T_2$, drawn uniformly at random (which corresponds to the distribution induced by the Haar measure over the group of unitary matrices). As in the previous section, we compute the singular value decomposition decomposition of $M = e^{i\alpha_1}T_1+e^{i\alpha_2}T_2 = U \Sigma V^\dagger$, where $\alpha_1$ and $\alpha_2$ are two parameters. Let $X$ and $Y$ be the right and left singular vectors associated with a singular value $\mu$. More precisely, $\mu$ is the $j$-th coefficient of $\Sigma$ for some $j$, and $X$ and $Y$ are respectively the $j$-th column of $V$ and $U$. We are interested in the correlation between $T_1X$ and $T_2X$, that is
\[C(T_1X, T_2X) = \frac{X^\dagger T_1^\dagger T_2 X}{||T_1X||_2\cdot ||T_2X||_2}\]
Because $T_1$, $T_2$, $U$ and $V$ are unitary matrices, we have $||T_1X||_2= ||T_2X||_2=1$. Thus, the correlation $C(T_1X, T_2X)$ is equal to the $i$-th diagonal value of the matrix $C = U^\dagger T_1^\dagger T_2 U$. As a product of unitary matrices, $C$ is unitary. We are now going to show that $C$ is also a diagonal matrix, which will imply that its diagonal coefficients are complex numbers of modulus 1. Using both the definition of $M$ and its singular value decomposition, one can write
\begin{align*}
M^\dagger M &= (e^{i\alpha_1}T_1+e^{i\alpha_2}T_2)^\dagger(e^{i\alpha_1}T_1+e^{i\alpha_2}T_2) = 2I+e^{i(\alpha_2-\alpha_1)}T_1^\dagger T_2 + e^{i(\alpha_1-\alpha_2)}T_2^\dagger T_1\\
M^\dagger M &= (V\Sigma U^\dagger)^\dagger(V\Sigma U^\dagger) = U\Sigma^2U^\dagger.
\end{align*}
Combining both equations gives $e^{i(\alpha_2-\alpha_1)}T_1^\dagger T_2 + e^{i(\alpha_1-\alpha_2)}T_2^\dagger T_1 = U\Sigma^2U^\dagger - 2I$, which implies that
\[(e^{i(\alpha_2-\alpha_1)}C)+(e^{i(\alpha_2-\alpha_1)}C)^\dagger = 
U^\dagger(e^{i(\alpha_2-\alpha_1)}T_1^\dagger T_2 + e^{i(\alpha_1-\alpha_2)}T_2^\dagger T_1)U = \Sigma^2 - 2I.\]
This shows that the real part of $Z = e^{i(\alpha_2-\alpha_1)} C = \Sigma^2/2-I$ is a diagonal matrix. Because $T_1$ and $T_2$ are random unitary matrices, then all coefficients from $Z$ are distinct with probability~1.
We are going to show that the imaginary part of $Z$ is also diagonal. First, observe that the matrices $(Z+Z^\dagger)$ and $(Z-Z^\dagger)$ commute:
$(Z+Z^\dagger)(Z-Z^\dagger) = Z^2-(Z^\dagger)^2 = (Z-Z^\dagger)(Z+Z^\dagger)$.
More precisely, this means that $\Re(Z)$ and $\Im(Z)$ commute, and that both matrices stabilize the eigen-spaces of the other matrix. If $\Re(Z)$ is a diagonal matrix with distinct coefficients, then each of its eigen-spaces has dimension 1 and is spanned by one of the vectors of the canonical basis. Thus, each vector of the canonical basis is an eigenvector of $\Im(Z)$, which in turn is diagonal. Hence, $Z$ is diagonal, and because it is also unitary then diagonal coefficients must be complex of modulus 1. 
\smallbreak
Going back to our correlations, we showed that the correlation between $T_1X$ and $T_2X$ is a complex of modulus 1, such that $\Re(e^{i(\alpha_2-\alpha_1)}C(T_1X, T_2X)) = \mu^2/2-1$. Thus, we have 
\begin{equation}
C(T_1X, T_2X) = e^{i(\alpha_1-\alpha_2)}\cdot(\mu^2/2-1\pm i\mu\sqrt{1-\mu^2/4})
\label{eq:SM_correlation_unitary}
\end{equation}

\subsection{Simulations}
\label{sec:SM_simulations}

To perform the simulations presented in the article, we used a simple approach modelling the TM by a random matrix~\cite{popoff2010measuring, devaud2021speckle}.
When quantitatively comparing simulations to experimental results, the remaining grain size after binning is added to the random matrix by convolving each output dimension with the adequate Gaussian.

\section{Comparison of the different operators}
\label{sec:SM_comparison_operators}

We present in this section simulation results that compare the correlation achievements obtained for different operators: the SVD of $T_1 + T_2$ (eigenvalue of $(T_1 + T_2)^{\dagger}(T_1 + T_2)$), the eigen decomposition of $T_1^{\dagger} T_2$, of $T_1^{\dagger} T_2 + T_2^{\dagger} T_1$ (its symmetric version) and $T_1^{\dagger} (T_2 - T_1)$ (equivalent for the Wigner-Smith approach where $(T_2 - T_1)$ corresponds to the derivative part \cite{ambichl2017super}).
We only consider the case of square non-unitary matrices.
A more complete comparison goes beyond the frame of this work.

\begin{figure}[h!]
    \includegraphics[width=0.9\columnwidth]{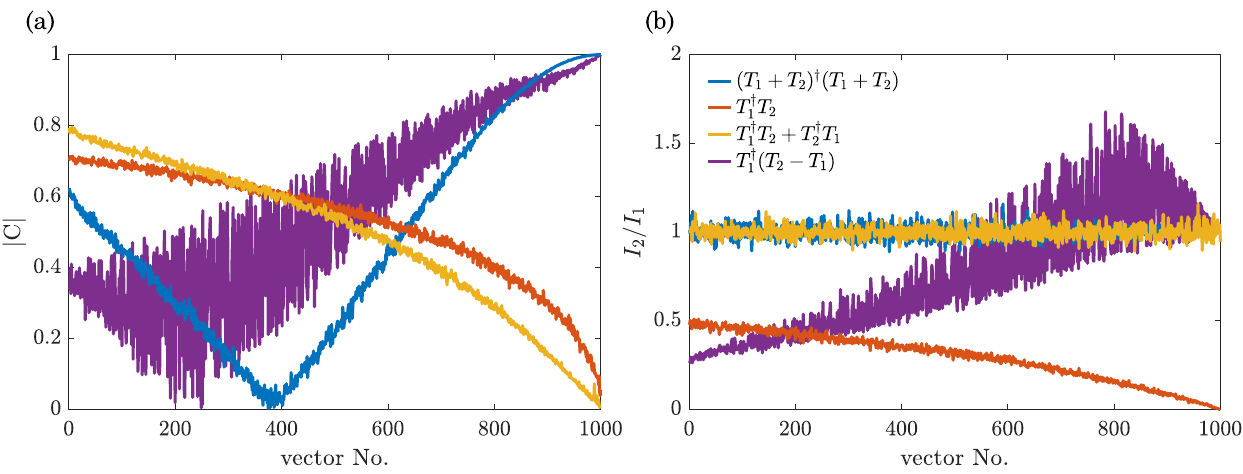}
    \phantomsubfloat{\label{fig:SM_comparison_methods-a}}
    \phantomsubfloat{\label{fig:SM_comparison_methods-b}}
    \caption{Comparison of different methods for correlating fields
    (a) The absolute value of the correlation is plotted as a function of the input vector number (ordered by decreasing the absolute value of the singular/eigenvalue).
    The simulation is performed with TMs of size $1000 \times 1000$.
    (b) Ratio of the total transmitted intensities for the same data as in (a).
    }
    \label{fig:SM_comparison_methods}
\end{figure}

For all four operators, the input vectors $X_s$ are sorted by decreasing the absolute value of the singular value/ eigenvalue (it is noteworthy that the eigenvalues are not necessarily real).
The resulting absolute values of the correlation (between $T_1 X_s$ and $T_2 X_s$) are plotted as a function of the number of the singular vector (No.  in \cref{fig:SM_comparison_methods-a}).

Only the SVD approach and the Wigner-Smith approach allow us to fully modulate the correlations ($|C| = 1$ in case of a square TM) while the other methods access a more restricted range with a lower maximum correlation observed at $0.8$ with the definition of \cref{eq:correlation_formula}.
However, it should be noted that perfect correlation is expected for the lowest singular value, which is the most sensitive to noise or phase-only modulation.
In the current experimental capabilities, the latter point diminishes the specific advantage of the SVD method, but continued progress in beam shaping control will overcome this purely experimental limitation.
Using the Wigner-Smith operator grants similar results to the SVD approach while losing the benefit of the sorting due to real singular values, which is of practical interest.
It also lacks easy and independent control of the absolute value and the phase of the correlation.

The methods of \cite{pai2021scattering} and its symmetrized version give similar correlation values. 
The main difference between the latter two methods is visible in the transmitted intensities presented in \cref{fig:SM_comparison_methods-b}.
The ratio of the total transmitted intensities is displayed and one can observe that all symmetrized versions lead to balanced intensities (i.e., the symmetrized version of \cite{pai2021scattering} and SVD method).
This observation is not surprising as the two TMs play non-symmetric roles.

\section{Extension to multimode fibres}

\begin{figure}[h!]
    \includegraphics[width=0.4\columnwidth]{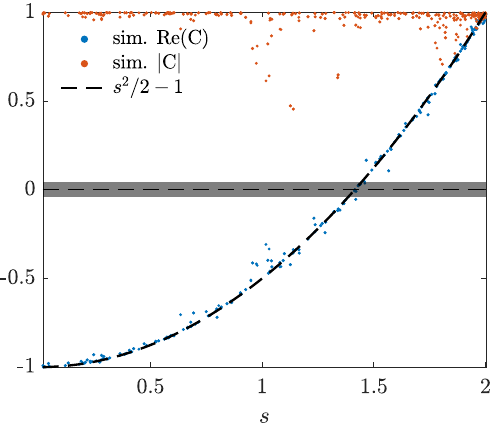}
    \caption{Simulation of field correlation for multimode fibers.
    Real part (blue dots) and absolute value (orange dots) of the correlations when summing two unitary TMs.
    The analytical prediction of~\cref{eq:SM_correlation_unitary} is presented by the black dashed line.
    The horizontal black dashed line represents the mean correlation for a random input vector and the grey-shaded area is its standard
    deviation.
    The data are not averaged over several realizations.
    }
    \label{fig:SM_fiber}
\end{figure}

We also performed simulations to predict the results of the technique for unitary systems.
A good example of a complex but lossless platform is multimode fibers (MMF).
In addition, the finite number of propagation modes makes it possible to measure complete TMs.
We performed the simulations using the code of~\cite{sebastien_m_popoff_2020_4075298} which solves the transverse scalar propagation equation and allows
to calculate the modes for fibers with arbitrary index profiles and their TMs.
We have applied it to measure the TMs of a \SI{10}{\centi \metre} long step-index MMF with $0.22$ numerical aperture and \SI{25}{\micro \metre} radius illuminated with a light of either \SI{700}{\nano \metre} or \SI{800}{\nano \metre}.
Due to the variation of the wavelength, the number of propagation modes is different.
However, the code allows them to be expressed on the pixel basis, as in the experiments, so that their size is the same, allowing them to be added together.
Thus, once the zero singular values corresponding to the rank difference between the two TMs are removed, the correlation results presented in~\cref{fig:SM_fiber} are obtained.

Two main points are worth noting: the absolute value of the correlation, equal to unity, does not depend on the singular vector and the real part of the correlation well follows the law predicted in~\cref{eq:SM_correlation_unitary}.
Thus even if the resulting fields are perfectly correlated, the unitarity of the transformation prevents any variations.
For this reason, losses and non-unitary transformations are often sought~\cite{nardi2021controlling}.

\end{document}